\documentclass[12pt]{article}
\usepackage{graphicx}
\usepackage[bookmarksnumbered,colorlinks,plainpages]{hyperref}

\begin{document}

\title{Applying incomplete statistics to nonextensive systems with different $q$ indices}

\author{L. Nivanen$^{\dag}$, M. Pezeril$^{\ddag}$, Q.A. Wang$^{\dag}$ and A. Le M\'ehaut\'e$^{\dag}$\\
$^{\dag}$Institut Sup\'erieur des Mat\'eriaux du Mans, \\
44, Avenue F.A. Bartholdi, 72000 Le Mans, France\\
$^{\ddag}$Laboratoire de Physique de l'\'etat Condens\'e, \\ Universit\'e du Maine,
72000 Le Mans, France\\}

\date{}

\maketitle

\begin{abstract}
The nonextensive statistics based on the $q$-entropy
$S_q=-\frac{\sum_{i=1}^v(p_i-p_i^q)}{1-q}$ has been so far applied to systems in
which the $q$ value is uniformly distributed. For the systems containing different
$q$'s, the applicability of the theory is still a matter of investigation. The
difficulty is that the class of systems to which the theory can be applied is
actually limited by the usual nonadditivity rule of entropy which is no more valid
when the systems contain non uniform distribution of $q$ values. In this paper,
within the framework of the so called incomplete information theory, we propose a
more general nonadditivity rule of entropy prescribed by the zeroth law of
thermodynamics. This new nonadditivity generalizes in a simple way the usual one and
can be proved to lead uniquely to the $q$-entropy.
\end{abstract}

PACS numbers

{05.70.Ln} {Nonequilibrium and irreversible thermodynamics}

{05.45.-a} {Nonlinear dynamics and nonlinear dynamical systems}

{05.20.-y} {Classical statistical mechanics}

\section{Introduction}
This work addresses the problem of the application of the nonextensive statistics
based on a $q$-entropy to systems containing non uniform distribution of $q$ values,
i.e., containing subsystems which have different $q$'s. The $q$-entropy for open
systems is given by
\begin{equation}                                    \label{1}
S_q(p_1,p_2,...p_v,q)=-\frac{\sum_{i=1}^v(p_i-p_i^q)}{1-q},\;\;\; q\geq 0
\end{equation}
where $p_i$ is the probability that the system is found at the state $i$, and $v$ is
the total number of states occupied by the system (let Boltzmann constant $k$ be
unity). This entropy is proposed first by Havrda, Charvat\cite{Esteban,Havrda} and
Daroczy\cite{Daroczy}, who deleted the additivity requirement from the entropy
axioms of Shannon\cite{Shannon}, as a possible generalization of Shannon entropy
$S_1=-\sum_{i=1}^vp_i\ln p_i$\footnote{In the formula of Havrda, Charvat and
Daroczy, the denominator was $2^{1-q}-1$ instead of $1-q$ in order that
$S_q(1/2,1/2,q))=1$ for whatever $q$.}. Daroczy\cite{Daroczy} further discussed in
detail the properties of this entropy, including its nonadditivity for a composite
system $A+B$ containing independent subsystems $A$ and $B$ :
\begin{equation}                                    \label{2}
S_q(A+B)=S_q(A)+S_q(B)+(1-q)S_q(A)S_q(B).
\end{equation}

In the last decades, a nonextensive statistical mechanics (NSM) is derived by
Tsallis and coworkers from maximizing $S_q$\cite{Tsal88,Tsal03a}. A main character
of NSM is the $q$-exponential probability distribution :
\begin{equation}                                    \label{3}
p_i=\frac{1}{Z_q}[1-a\beta E_i]^{1/a} \;\;\; [\cdot]>0.
\end{equation}
where $a$ equals $1-q$ or $q-1$ according to the maximization process of
$S_q$\cite{Wang02c} and $E_i$ can be the energy (or any other thermodynamic quantity
introduced into the process of maximum entropy) of the system at the state $i$. The
validity range of NSM is still under investigation. Actually, a point of view is
widely accepted that NSM can be used to describe nonequilibrium systems at
stationary state whose probability distribution is time independent. This position
is supported by the fact that all the actual applications of NSM and its
$q$-exponential distribution in Eq.(\ref{3}) are related to critical, nonequilibrium
and chaotic systems.

In this paper, we will focalize our attention on an important question about the
applicability of NSM to nonextensive systems containing subsystems with different
$q$'s. For this purpose, we take it for granted that : 1) the $q$-entropy, as an
information or uncertainty measure with respect to the probability $p_i$, is valid
for some nonequilibrium nonextensive systems in steady or stationary states with
time independent probability distribution $p_i$; 2) the Janyes' inference method of
maximum entropy\cite{Jaynes,Tribus} applies for these states, i.e., the time
independent probability distribution on these states can be deduced from maximizing
the $q$-entropy.

The above assumptions are necessary for NSM to be applied to nonequilibrium systems
with the distribution given by Eq.(\ref{3}). Due to the maximum $q$-entropy, the
framework of the conventional equilibrium thermodynamics (CET) can be copied for
stationary nonequilibrium systems. Essential for this work is the establishment of
the zeroth law or the equality between the intensive variable $\beta$ at state
maximizing entropy $S_q$. It is worth noticing that, due to the different formalisms
of NSM proposed in the past 15 years, the intensive parameter $\beta$ has several
definitions and interpretations which sometimes cause confusion. A comment on this
subject was given in ref.\cite{Wang04a}. In this work, we choose a formalism
allowing $\beta=\frac{\partial S_q}{\partial U_q}$ (we will call it {\it inverse
temperature} from now on), here $U_q$ is the expectation of $E_i$ discussed below.

The application of NSM to systems without uniform $q$ value distribution is
fundamental because a large number of systems in Nature, especially the systems far
from equilibrium, the chaotic and fractal systems, are inhomogeneous and may be
divided into subsystems with different nonextensive property and $q$ values. If NSM
does not allow treatments of such systems, there would be an handicap in its
theoretical formulation. In other words, if NSM can separately treat $A$ and $B$
each having its own $q$, it is also expected to be valid for the total system $A+B$
and to provide method to derive $q_{A+B}$ from $q_A$ and $q_B$. This problem was
recently discussed by many scientists\cite{Sasaki,Abe,Tsal03} on the basis of
Eq.(\ref{2}). The attempts were interesting but a rigorous mathematical formulation
of zeroth law relating $\beta(A)$ and $\beta(B)$ is still missing. This result is
not surprising because Eq.(\ref{2}) is only the nonadditivity rule for the systems
containing subsystems with same $q$. Its application to different $q$-systems is in
fact forbidden.

The aim of this paper is to show that, by using a more general nonextensive rule of
entropy which generalizes Eq.(\ref{2}), the establishment of zeroth law relating
$\beta(A)$ and $\beta(B)$ is possible for the systems having different $q$'s. This
approach was recently described within the formalism of NSM based on the usual
normalized probability\cite{Wang03e}. In this paper, we are interested in extending
the approach to the formalism of incomplete statistics proposed several years ago by
one of us\cite{Wang02c,Wang01a,Wang02b}, and providing in addition a detailed
description of the calculus allowing one to determine the $q$ of composite systems
from the $q$'s of the subsystems.

Incomplete statistics is a version of NSM based on the notion of {\it incomplete
information} associated with the normalization
\begin{equation}                                    \label{x1}
\sum_{i=1}^wp_i^q=1\;\;\;(q\geq 0),
\end{equation}
here $w$ is only the number of states which are countable or accessible to our
treatments and may be larger or smaller than the number $v$ of the physical states
occupied by the systems\cite{Wang02b,Wang04}. It is proved\cite{Wang02b,Wang04} that
this incomplete normalization may arise for the systems whose phase space is
occupied in heterogeneous (hierarchical or fractal) way and coarse grained. An
example of such systems is the polymers described in coarse graining way by using
contact matrix which makes it impossible to completely calculate the information
contained in the sequence of monomers units\cite{Lebowitz}. It has been
shown\cite{Wang02b,Wang04} that $q=d_f/d$ for the systems whose phase space volume
is simple fractal of dimension $d_f$, where $d\geq 1$ is the dimension of the phase
space when it is smoothly occupied. In this case, $S_q$ can be recast into
$S_q=\frac{1-\sum_{i=1}^wp_i}{1-q}$\cite{Wang01a}. For a composite system having
joint probability given by the product of the probabilities of its subsystems, i.e.,
$p_{ij}(A+B)=p_i(A)p_j(B)$, we still have Eq.(\ref{2}) with only the change $(1-q)$
to $(q-1)$\cite{Wang02c}.

\section{A nonadditivity of entropy prescribed by zeroth law}
Eq.(\ref{2}) has been shown to be a special case of the composition
rule\cite{Abe02a}
\begin{equation}                                    \label{5}
H[Q(A+B)]=H[Q(A)]+H[Q(B)]+\lambda_Q H[Q(A)]H[Q(B)],
\end{equation}
where $H[Q]$ is certain differentiable function satisfying $H[0]=0$, $\lambda_Q$ is
a constant, and $Q$ is either $S_q$ or $U_q$\cite{Wang02a}, which allows the
establishment of zeroth law of thermodynamics in nonextensive systems. As a matter
of fact, Eq.(\ref{5}) has been established\cite{Abe02a,Wang02a} for the class of
systems containing only subsystems having the same $q$. It must be generalized for
the systems whose subsystems have different $q$'s. This generalization is
straightforward if we replace the Eq.(1) of reference \cite{Abe02a}, i.e.,
$S(A+B)=f\{S(A),S(B)\}$ for uniform $q$, by
$H_{q}[S_q(A+B)]=f\{H_{q_A}[S_{q_A}(A)],H_{q_B}[S_{q_B}(B)]\}$ (or by
$H_{q}[S_q(A+B)]=H_{q_A}[S_{q_A}(A)]+H_{q_B}[S_{q_B}(B)]
+g\{H_{q_A}[S_{q_A}(A)],H_{q_B}[S_{q_B}(B)]\}$) where $H_q(S_q)$ is a functional
depending on $q$'s with the same form for the composite system as for the
subsystems, where $q$, $q_A$ and $q_B$ are the parameters of the composite system
$A+B$, the subsystems A and B, respectively. This functional $H_q(S_q)$ is necessary
for the equality to hold in view of the different $q$'s in the entropies of
different subsystems. The function $f$ (or $g$) is to be determined by the zeroth
law. Now repeating the mathematical treatments described in the references
\cite{Abe02a,Wang02a}, we find
\begin{equation}                                    \label{5x}
H_q[Q(A+B)]=H_{q_A}[Q(A)]+H_{q_B}[Q(B)]+\lambda_Q H_{q_A}[Q(A)]H_{q_B}[Q(B)].
\end{equation}
Eq.(\ref{5}) turns out to be a special case of Eq.(\ref{5x}), and Eq.(\ref{2})
corresponds to a $H_q[S_q]$ which is identity function. Now for $A$ and $B$ each
having its own $q$, in view of the fact that the $q$-entropy of Eq.(\ref{1}) must
have the same form for any system, a simple choice is $H_q[S_q]=(q-1)S_q$ for the
version of $S_q$ within incomplete statistics, which means
\begin{eqnarray}                                    \label{2a}
(q-1)S_q(A+B) &=& (q_{A}-1)S_{q_A}(A)+(q_{B}-1)S_{q_B}(B) \\ \nonumber &+&
\lambda_S(q_{A}-1)(q_{B}-1)S_{q_A}(A)S_{q_B}(B).
\end{eqnarray}
This nonadditivity generalizes Eq.(\ref{2}) which is recovered when $q=q_A=q_B$.

\section{Uniqueness of $S_q$}
Eq.(\ref{2}) has been shown to lead uniquely to the $q$-entropy of
Eq.(\ref{1})\cite{Santos,Abe00x} when the product joint probability holds. This
result is not complete because it was obtained only for the systems of uniform $q$
value distribution. The reasoning can also be straightforwardly generalized to
inhomogeneous systems containing different $q$'s. In order to do this, the axiom
[II]* of the reference \cite{Abe00x}, i.e., Eq.(\ref{2}), should be replaced by
Eq.(\ref{2a}), with other axioms unchanged. In this way, the Eq.(16) of
\cite{Abe00x}, i.e., $L_q(r^m)=\frac{1}{(1-q)}\{[1+(1-q)L_q(r)]^m-1\}$ should be
replaced by $[1+(q-1)L_q(r^m)]=\prod_{l=1}^m[1+(q_l-1)L_{q_l}(r)]$ derived from
Eq.(\ref{2a}), where $L_q$ is the functional of the $q$-entropy, $q$ and $q_l$ are
respectively the parameter of the composite system and of the subsystems labelled by
$l$ ($l=1,2,...m$), and $r$ the number of states in each subsystem. This
relationship can be written as
$\ln[1+(q-1)L_q(r^m)]=\sum_{l=1}^m\ln[1+(q_l-1)L_{q_l}(r)]$. The same mathematics as
in \cite{Santos,Abe00x} leads to
$S_{q_l}(1/r,q_l)=L_{q_l}(r)=\frac{r^{q_l-1}-1}{q_l-1}$ and
$S_{q_l}(p_1,p_2,...p_v,q_l)=\frac{\sum_{i=1}^wp_i-1}{q_l-1}$.

This uniqueness of $S_q$ implies that Eq.(\ref{2a}) is intrinsically a possible
composition rules of the $q$-entropy which becomes in this way invariant with
respect to the inhomogeneity of $q$ values in the systems of interest. The $q$ value
in the entropy of the composite system can be determined from the $q$'s of the
subsystems, which will be described below.

\section{A zeroth law}

Now from Eq.(\ref{2a}), according to our starting assumption that the stationary
state of the composite system $A+B$ maximizes its $q$-entropy, i.e., $dS(A+B)=0$, we
get :
\begin{eqnarray}                                    \label{8}
\frac{(q_{A}-1)dS(A)}{1+(q_{A}-1)S(A)}+\frac{(q_{B}-1)dS(B)}{1+(q_{B}-1)S(B)}=0
\end{eqnarray}
or $\frac{(q_{A}-1)dS(A)}{\sum_{i}p_i(A)}+\frac{(q_{B}-1)dS(B)}{\sum_{i}p_i(B)}=0$.

As to the nonadditivity of the quantity $U_q$, it is in fact uniquely determined by
the product joint probability $p_{ij}(A+B)=p_i(A)p_j(B)$ which implies that, if we
want to split the thermodynamics of the composite system into those of the
subsystems, the expectation should be defined by $U_q=\sum_ip_iE_i$ or
$U'_q=\sum_ip_iE_i/\sum_ip_i$ (it is evident that the normalized expectation
$U_q=\sum_ip_i^qE_i$ is no more factorizable). It is easy to show that, using $U_q$
and Eq.(\ref{x1}) as constraints, the maximization of $S_q$ gives :
\begin{equation}                                    \label{4}
p_i=\frac{1}{Z_q}[1-(q-1)\beta E_i]^{1/q-1} \;\;\; [\cdot]>0
\end{equation}
where the partition function is given by $Z_q=\{\sum_i[1-(q-1)\beta
E_i]^{q/q-1}\}^{1/q}$. With some mathematics, we find
\begin{equation}                                    \label{4x}
\sum_ip_i=Z^{q-1}+(q-1)\beta U_q.
\end{equation}
Then combining the product probability and Eq.(\ref{4x}), we get
\begin{eqnarray}                                    \label{15}
& Z_q^{1-q}(A+B)+(q-1)\beta(A+B) U_q(A+B) \\\nonumber & =
[Z_{q_A}^{1-q_A}(A)+(q_{A}-1)\beta(A)
U_{q_A}(A)][Z_{q_B}^{1-q_B}(B)+(q_{B}-1)\beta(B) U_{q_B}(B)].
\end{eqnarray}
The energy conservation law of the total system $dU_q(A+B)=0$ leads to
\begin{eqnarray}                                    \label{15xx}
\frac{(q_{A}-1)\beta(A)dU_{q_A}(A)}{\sum_ip_i(A)}
+\frac{(q_{B}-1)\beta(B)dU_{q_B}(B)}{\sum_ip_i(B)}=0
\end{eqnarray}
which suggests the following nonadditivity :
\begin{eqnarray}                                    \label{x14}
\frac{(q_{A}-1)dU_{q_A}(A)}{\sum_ip_i(A)}+\frac{(q_{B}-1)dU_{q_B}(B)}{\sum_ip_i(B)}=0.
\end{eqnarray}
This relationship should be considered as a generalization of the additivity rule
$dU(A)+dU(B)=0$ of CET. From Eq.(\ref{8}) to Eq.(\ref{x14}), we see the necessity to
choose the unnormalized expectation in order to split the thermodynamics of the
composite system into those of the subsystems via the energy nonadditivity given by
Eq.(\ref{x14}). By splitting thermodynamics, we means that the thermodynamics of the
subsystems can be formulated exactly in the same way and with the same mathematical
definition of all the thermodynamic variables and functions as for the total system.
Without this splitting, the establishment of zeroth law would be impossible.

Now comparing Eq.(\ref{8}) and Eq.(\ref{x14}), we obtain
\begin{eqnarray}                                    \label{16}
\beta(A)=\beta(B)
\end{eqnarray}
with $\beta(A)=\frac{\partial S_{q_A}(A)}{\partial U_{q_A}(A)}$ and
$\beta(B)=\frac{\partial S_{q_B}(B)}{\partial U_{q_B}(B)}$ which can be
straightforwardly derived from Eq.(\ref{1}) and Eq.(\ref{4x}). So this zeroth law is
independent of whether or not the subsystems have the same $q$.

\section{How to determine the composite $q$}
If $q_A$, $q_B$, $p_i(A)$ and $p_j(B)$ of the subsystems are well known, the
parameter $q$ of the composite system is uniquely determined by the product joint
probability. Using the incomplete normalization of the joint probability, we obtain
:
\begin{eqnarray}                                    \label{x7}
\sum_{i=1}^{w_A}p_i^q(A)\sum_{i=1}^{w_B}p_i^q(B)=1
\end{eqnarray}
or
\begin{eqnarray}                                    \label{7}
\sum_{i=1}^{w_A}(p_i^{q_A})^{q/q_A}(A)\sum_{i=1}^{w_B}(p_i^{q_B})^{q/q_B}(B)=1
\end{eqnarray}
which means $q_A<q<q_B$ if $q_A<q_B$ and $q=q_A=q_B$ if $q_A=q_B$. In what follows,
the above result will be generalized to more complicated composite system containing
$N>2$ subsystems each having its own $q$. Suppose an integer ensemble
$\aleph_m=\verb"N"\cap[1,m]=\{1,2,...,m\}$. The subsystems are characterized by
$[p_{i_k}, q_k, n_k]$ where $k\in\aleph_N$ and $i_k\in\aleph_{n_k}$. The joint
probability is given by
\begin{eqnarray}                                    \label{23}
p_{i_1,i_2...i_N}=\prod_{k=1}^Np_{i_k}
\end{eqnarray}
where
\begin{eqnarray}                                    \label{24}
\sum_{i_k=1}^{n_k}p_{i_k}^{q_k}=1   \;\;\; \forall k\in\aleph_N.
\end{eqnarray}
Suppose that there is a $q$ such that the joint probability is normalized as follows
\begin{eqnarray}                                    \label{25}
\sum_{i_1,i_2...i_N=1}^{n_1n_2...n_N}p_{i_1,i_2...i_N}^q=1,
\end{eqnarray}
then $q$ is determined by
\begin{eqnarray}                                    \label{26}
\prod_{k=1}^N\sum_{i_k=1}^{n_k}p_{i_k}^q=1.
\end{eqnarray}

To see how to estimate $q$, let us first suppose that each subsystem has
equiprobable states. One gets, for the $k^{th}$ subsystem, $p_k=(1/n_k)^{1/q_k}$.
From Eqs.(\ref{23}) to (\ref{26}), we obtain
\begin{eqnarray}                                    \label{27}
(\prod_{k=1}^Nn_k)^{1/q}=\prod_{k=1}^N(n_k^{1/q_k}).
\end{eqnarray}

Property 1 : $q\in [\min{q_k},\max{q_k}]\;\;\;\forall k\in\aleph_N$.

Proof : Eq.(\ref{27}) can be recast into $(1/q)\sum_k\ln n_k=\sum_k(1/q_k)\ln n_k$
or $(1/q)=[\sum_k(1/q_k)\ln n_k]/[\sum_k\ln n_k]$. So $1/q$ can be seen as a
barycenter of the terms $1/q_k\geq 0$ with the coefficients $\ln n_k \geq 0$. Hence
we have $q\in [\min{q_k},\max{q_k}]$. The property 1 is proved.

We see that, if all the subsystems have the same number of states, we get
\begin{eqnarray}                                    \label{28}
(1/q)=\frac{1}{N}\sum_{k=1}^N(1/q_k).
\end{eqnarray}

Property 2 : The property 1 can be generalized to subsystems having non-equiprobable
states.

Proof : Suppose $q>\max q_k \;\;\;\forall k\in\aleph_N$, we have
$\sum_{i_k}^{n_k}p_{i_k}^q<\sum_{i_k}^{n_k}p_{i_k}^{q_k}=1$ and, as a consequence,
$\prod_k\sum_{i_k}^{n_k}p_{i_k}^q=\sum_{i_1,i_2...i_N}^{n_1n_2...n_N}p_{i_1,i_2...i_N}^q<1$
which is at variance with the normalization of the joint probability. So $q\leq \max
q_k \;\;\;\forall k\in\aleph_N$. In the same way, it can be shown that $q\geq \min
q_k$. Property 2 is proved.

\section{Conclusion}
A general entropy composition rule prescribed by the zeroth law of thermodynamics is
proposed for the systems containing subsystems with different $q$'s. The usual
entropy composition rule of NSM can be recovered when $q$ is the same for all
subsystems. The zeroth law, or the equality between the intensive parameters $\beta$
of each subsystems, is proved to be independent of the $q$ values of subsystems. It
is shown that $q$-entropy $S_q$ is unique in the context of this general
nonadditivity. A calculus is provided for deriving the $q$ value characterizing the
composite system from the $q$ values and the probability distributions of
subsystems.

It is worth mentioning that this work is carried out within incomplete statistics by
using the unnormalized $q$-expectation. It is found that, if one wants to establish
relations between $\beta(A)$ and $\beta(B)$ when the subsystems have different
$q$'s, the usual expectation $U=\sum_ip_i^qE_i$ for incomplete distribution cannot
be used due to the factorization of joint probability $p_{ij}(A+B)=p_i(A)p_i(B)$. We
would like to mention in passing that this constraint of thermodynamic splitting and
of zeroth law establishing for inhomogeneous $q$-systems fundamentally changes the
formulation of NSM with normalized distribution $\sum_i p_i=1$ because in this case
we have to use a general product rule $p_{ij}^q(A+B)=p_i^{q_A}(A)p_i^{q_B}(B)$ for
``formally independent''\footnote{Product joint probability means independence of
the subsystems and no interaction in the conventional probability theory. This
meaning does not apply for nonextensive systems having interacting and correlating
subsystems due to which the nonextensivity of entropy and energy can arise. So by
``formal independence'' we mean a statistical situation where interacting systems
verify the product rule of joint probability.} subsystems $A$ and $B$ instead of the
usual product probability $p_{ij}(A+B)=p_i(A)p_i(B)$ as used in this work. Detailed
description of this approach can be found in \cite{Wang03e}.

\end{document}